# Reaction front development from ignition spots in *n*-heptane/air mixtures: low-temperature chemistry effects induced by ultrafine water droplet evaporation

Zhou Yu and Huangwei Zhang*

Department of Mechanical Engineering, National University of Singapore, 9 Engineering Drive 1, Singapore 117576

**Abstract**

Effects of low-temperature chemistry induced by ultrafine water droplet evaporation on reaction front development from an ignition spot with temperature gradient are studied in this work. The Eulerian−Eulerian method is used to simulate the gas−liquid two-phase reactive flows and the physical model is one-dimensional spherical reactor with stoichiometric gaseous *n*-heptane/air mixture and ultrafine monodisperse water droplets (initial diameter 5 micrometres). Homogeneous ignitions of two-phase mixtures are first simulated. The water droplets can complete evaporation in the reactor prior to ignition, and hence pronouncedly reduce gas temperature, which may induce the low-chemistry reactions. It is found that the turnover temperature for negative temperature coefficient range increases with droplet volume fraction. Three-stage ignitions are present when the volume fraction is beyond a critical value, i.e., low-temperature, intermediate-temperature, and high-temperature ignitions. The chemical explosive mode analysis also confirms the low-chemistry reactions induced by the evaporation of ultrafine water droplets. Then reaction front development from an ignition spot with temperature gradient in two-phase mixtures is analysed based on one-dimensional simulations. Different modes for reaction front origin in the spot are identified, based on the initial gas temperature and lower turnover temperature. Specifically, the reaction front can be initiated at the left and right ends of the ignition spot, and inside it. Detailed reaction front developments corresponding to the above three modes are discussed. Besides, the pressure wave from high-temperature ignition is important, compared to those from low- and intermediate-chemistries. The reaction front propagation speed and thermal states of fluid particles corresponding to different reaction front initiation modes are analysed. Moreover, autoignition modes are summarized in the diagrams of normalized temperature gradient versus normal acoustic time and droplet volume fraction. The detonation limits of two-phase mixtures highly depend on the droplet volume fraction and are not regularly peninsular-shaped, like those for purely gaseous mixtures.

---

* Corresponding author. E-mail: huangwei.zhang@nus.edu.sg.

# 1 Introduction

Downsizing of spark-ignition engines with turbocharging technology is promising, since it is deemed a novel solution for pollutant reduction and fuel economy [1]. However, knocking combustion is likely to happen [2–5], caused by the interactions between acoustic wave and chemical reaction when end gas autoignites [1]. In particular, reaction wave and detonation development subject to localized reactivity non-uniformity (e.g. temperature gradient) in the chamber plays a dominant role in inducing this hazardous phenomenon [6,7].

Zel'dovich [8] identified different autoignition modes caused by a hot spot with thermal inhomogeneity, i.e., subsonic reaction wave, detonation development and supersonic reaction wave. Bradley and his co-workers [9–12] further introduced a detonation peninsula (termed as *Bradley's diagram* hereafter), parameterized by normalized temperature gradient $\xi$ and normalized acoustic time $\varepsilon$. After that, numerous simulations have been performed to uncover the underpinning mechanism of autoignition and detonation development from a localized ignition spot [13–27]. For instance, Dai et al. [20–24] investigated the various effects on autoignition and detonation development in DME/air and $n$-$C_7H_{16}$/air mixtures under engine-relevant conditions. It is noted that multi-stage ignition occurs at low initial temperature for large hydrocarbon fuels. In Ref. [21], Dai and Chen found that the temperature gradient of the hot spot is able to affect the interactions between multi-stage ignition and pressure waves. Pan et al. [15] studied the role of low temperature chemistry in combustion mode development and autoignition position. Moreover, Terashima et al. [17–19] unveiled the mechanisms of pressure wave development in end-gas autoignition during knocking combustion. They also found that both the amplitude of pressure oscillations and timing of knocking occurrence are affected by low-temperature chemistry and strong pressure wave is induced by a hot spot with high reactivity. Besides the above one-dimensional simulations, the effects of inhomogeneities of thermochemical conditions (e.g., temperature or composition) on autoignition and knock formation were also investigated by Luong et al. [28,29] with multi-dimensional simulations. It is shown that reduction of energetic length scale would be helpful for mitigating knocking propensity. They introduced two parameters to predict the knocking intensity, i.e., detonation propensity and heat release rate fraction, which show good correlations with the knock intensity when they are plotted against the normalized acoustic time $\varepsilon$.

It is well known that water injection technology is an effective approach to mitigate or alleviate knock in IC engines [30–33]. This is because evaporation of liquid water can reduce the in-cylinder temperature, because of high latent heat of vaporization and specific heat capacity of water vapour [30,34]. This technology has been vigorously studied in recent years. For instance, Wang et al. [33] investigated the possibility of injected water to extend the knock limits of a spark-ignition engine fuelled with kerosene. They found that the knock limit of their engine is significantly extended via water injection. Besides, Miganakallu et al. [35] studied the effects of liquid water/methanol injection on engine borderline knock conditions. They observed that addition of water in the fuel blends promotes combustion stability and considerably reduce the gas temperature. Zhuang et al. [36] also investigated the benefits of water injection on downsized boosted SI engine and pointed out that the water injection can effectively reduce the NO and CO emissions with increased injected water percentage.

Numerical simulations on the effects of water sprays on detonation development relevant to internal combustion engine conditions are also available. Zhuang et al. [37] studied autoignition and detonation development due to a hot spot in $n$-$C_7H_{16}$/air mixtures with liquid water droplets. The influences of droplet diameter and number density on reactive front development were discussed. However, detonation development regime associated with thermochemical properties of the hot spot (e.g., excitation time and acoustic time) in water-containing mixtures was not studied therein. More recently, the effects of water steam dilution on autoignition and detonation development induced by ignition spot with thermal inhomogeneity in *n*-C$_7$H$_{16}$/air mixture was numerically investigated in our previous work [38]. However, *in-situ* water droplet evaporation during reaction front development is not considered, and therefore how it affects initiation of the chemical reactions (such as low-temperature chemistry) of complex hydrocarbon fuels (e.g., *n*-heptane) from the ignition spot is not clear.

In this study, detailed numerical simulations of reaction front development from an ignition spot with temperature gradient in two-phase medium will be conducted. The physical model is one-dimensional spherical reactor filled with stoichiometric *n*-C$_7$H$_{16}$/air gas and ultrafine water droplets. The research objectives are: (i) to study the low-temperature chemistry effects (caused by the ultrafine water droplet evaporative cooling) on reaction front development, (ii) to identify the reaction front initiation mode subject to different droplet and gas properties, and (iii) to discuss the applicability of Bradley's diagram for the studied two-phase mixtures. The rest of the paper is structured as below. Sections 2 and 3 introduce the mathematical and physical models and data analysis method. Homogeneous ignition in two-phase mixtures is analysed in section 4. One-dimensional simulations of autoignition and detonation development due to ignition spot are studied in section 5. Finally, section 6 summarizes the main conclusions.

## 2 Mathematical and physical models
### 2.1 Gas phase equation

The governing equations of momentum, energy, and species mass fraction are solved for one-dimensional unsteady, multi-component, reacting flows. They can be written in a spherical coordinate as

$$\frac{\partial U}{\partial t} + \frac{\partial F(U)}{\partial r} + 2\frac{G(U)}{r} = F_v(U) + S_R + S_L, \qquad (1)$$

where $t$ and $r$ are time and radial coordinate, respectively. The vectors $U$, $F(U)$, $F_v(U)$, $G(U)$, $S_R$ and $S_L$ respectively have the following expressions

$$U = \begin{bmatrix} \rho Y_1 \\ \rho Y_2 \\ . \\ . \\ . \\ \rho Y_n \\ \rho u \\ E \end{bmatrix}, F(U) = \begin{bmatrix} \rho u Y_1 \\ \rho u Y_2 \\ . \\ . \\ . \\ \rho u Y_n \\ \rho u^2 + P \\ (E+P)u \end{bmatrix}, G(U) = \begin{bmatrix} \rho u Y_1 \\ \rho u Y_2 \\ . \\ . \\ . \\ \rho u Y_n \\ \rho u^2 \\ (E+P)u \end{bmatrix},$$

$$F_v(U) = \begin{bmatrix} -r^{-2}(r^2 \rho Y_1 V_1')_r \\ -r^{-2}(r^2 \rho Y_2 V_2')_r \\ . \\ . \\ . \\ -r^{-2}(r^2 \rho Y_n V_n')_r \\ r^{-2}(r^2 \tau_1)_r - 2\tau_2/r \\ r^{-2} q_r + \Phi \end{bmatrix}, S_R = \begin{bmatrix} \omega_1 \\ \omega_2 \\ . \\ . \\ . \\ \omega_n \\ 0 \\ 0 \end{bmatrix}, S_L = \begin{bmatrix} S_{m,1} \\ S_{m,2} \\ . \\ . \\ . \\ S_{m,n} \\ S_v \\ S_e \end{bmatrix}.$$

(2)

$\rho$ is the density and $u$ is the radial velocity. $E \equiv -P + \rho u^2/2 + \rho h$ is the total energy, with $h$ being the total enthalpy. $P$ is the pressure, obtained from the ideal gas equation of state $P = \rho RT/\bar{M}$. $R$ is the universal gas constant, $T$ and $\bar{M}$ are the temperature and mean molecular weight of the gaseous mixture, respectively. $Y_i$ and $\omega_i$ are the mass fraction and chemical reaction rate of $i$-th species, respectively. $n$ is the total number of species. The diffusion velocity $V_i'$ is determined using the mixture-averaged method. The chemical reaction rate $\omega_i$, thermodynamic and transport properties are calculated by CHEMKIN or TRANSPORT packages [39,40]. In Eq. (2), the subscript "$r$" in $F_v(U)$ stands for the partial derivative with respect to the spatial coordinate $r$. $\tau_1$ and $\tau_2$ are the viscous stresses, $q_r$ is the heat flux. Besides, $\Phi$ is the viscous dissipation rate. More details about the equations can be found in Refs. [24,41]. The effects of water droplets on gaseous phase are taken into consideration through the source/sink terms $S_L$, and their expressions will be given in Eq. (19).

## 2.2 Liquid phase equation

The Eulerian approach is applied to describe the liquid droplet phase. Similar approach is also used by Sanjosé et al. [42], Qiao et al. [43], as well as Eidelman and Burcat [44,45] for gas−liquid and gas−solid two-phase flows. In this study, the water droplet is assumed to be spherical. The droplet temperature is uniform due to the approximation of droplet infinite thermal conductivity [46,47]. The droplet breakup and deformation are not considered due to the small droplet diameters. The evolution of droplet diameter is governed by

$$\frac{\partial d}{\partial t} + u_d \frac{\partial d}{\partial r} = -\frac{2\dot{m}}{\pi \rho_d d^2}, \qquad (3)$$

where $d$, $u_d$ and $\rho_d$ are the droplet diameter, velocity, and material density, respectively. The evaporation rate $\dot{m}$ is modelled as [48]

$$\dot{m} = \pi d \rho D_{water,m} Sh \ln(1 + B_M), \qquad (4)$$

where $D_{water,m}$ is the binary diffusion coefficient of water vapour in the gaseous mixture and approximated following Ref. [49]. The Sherwood number $Sh$ is modelled as [48]

$$Sh = 2.0 + \frac{1}{F(B_M)} \left[ (1 + Re_d Sc)^{1/3} \max(1, Re_d)^{0.077} - 1 \right], \qquad (5)$$

where $F(B) = \frac{\ln(1+B)}{B}(1+B)^{0.7}$ is used to model the change of film thickness due to Stefan flow effects [48]. The droplet Reynolds number $Re_d$ is defined as

$$Re_d = \frac{\rho d |u - u_d|}{\mu}. \qquad (6)$$

In Eq. (5), the Schmidt number $Sc$ is estimated from

$$Sc = \frac{\mu}{\rho D_{water,m}}. \tag{7}$$

In Eqs. (4) and (5), $B_M$ is the Spalding mass transfer number

$$B_M = \frac{Y_{ds} - Y_{d\infty}}{1 - Y_{ds}}, \tag{8}$$

where $Y_{d\infty}$ is the water vapour mass fraction in the bulk gas. $Y_{ds}$ is the water vapour mass fraction at the droplet surface

$$Y_{ds} = \frac{W_{H_2O} X_{ds}}{W_{H_2O} X_{ds} + (1 - X_{ds})\overline{W}}. \tag{9}$$

Here $W_{H_2O}$ is the water molecular weight and $\overline{W}$ is that of the gas-phase mixture (excluding H₂O vapour). $X_{ds}$ is the water vapour mole fraction at the droplet surface

$$X_{ds} = \frac{P_{ref}}{P} \exp\left[\frac{L_v(T_{ref})}{R}\left(\frac{1}{T_{ref}} - \frac{1}{T_d}\right)\right]. \tag{10}$$

For water, the reference pressure is $P_{ref} = 1$ atm, the reference temperature is $T_{ref} = 370$ K, and the latent heat of vaporization is $L_v(T_{ref}) = 2,260$ J/g. It is noted that $T_{ref}$ is the corresponding boiling temperature under the reference pressure $P_{ref}$.

The equation of droplet velocity takes the following form

$$\frac{\partial u_d}{\partial t} + u_d \frac{\partial u_d}{\partial r} = \frac{F_s}{m_d}. \tag{11}$$

Note that only drag force $F_s$ is considered in our work and it is modelled using Schiller and Naumann's correlation [50], i.e., $F_s = \frac{m_d}{\tau_r} \cdot (u - u_d)$. $m_d = \rho_d \pi d^3 / 6$ is the mass of a single droplet. $\tau_r$ is the droplet momentum relaxation time and can be determined from [50]

$$\tau_r = \frac{\rho_d d^2}{18\mu} \frac{24}{C_d Re_d},\tag{12}$$

where $C_d$ is the drag coefficient [50]

$$C_d = \begin{cases} \frac{24}{Re_d}\left(1 + \frac{1}{6}Re_d^{2/3}\right), & if\ Re_d \leq 1000 \\ 0.44, & if\ Re_d > 1000 \end{cases}.\tag{13}$$

The studies by Cheatham and Kailasanath [51] confirm that Eq. (13) can accurately predict the gas velocity distributions in compressible two-phase flows.

The equation of droplet temperature reads

$$m_d C_{P,d}\left(\frac{\partial T_d}{\partial t} + u_d \frac{\partial T_d}{\partial r}\right) = h_c A_d (T - T_d) - \dot{m} L_v(T_d),\tag{14}$$

where $C_{P,d}$ is the constant pressure specific heat of the liquid phase and $A_d$ is the surface aera of a single droplet. $L_v(T_d)$ is the latent heat of vaporization at the droplet temperature [52]

$$L_v(T_d) = d_1 \cdot (1 - T_r)^{[(d_2 \cdot T_r + d_3) \cdot T_r + d_4] \cdot T_r + d_5},\tag{15}$$

where $d_1, d_2, d_3, d_4$, and $d_5$ are species-specific constants [52]. $T_r$ is defined as $T_r = T_d/T_{cr}$, where $T_{cr}$ is the critical temperature and is 647 K for water.

The convective heat transfer coefficient $h_c$ is

$$h_c = \frac{Nu k_g}{d},\tag{16}$$

where $k_g$ is the thermal diffusivity of gas phase. $Nu$ is the Nusselt number and can be estimated with Rans and Marshall model [53]

$$Nu = 2.0 + \frac{1}{F(B_T)}\left[(1 + Re_d Pr)^{1/3} \max(1, Re_d)^{0.077} - 1\right],\tag{17}$$

where $Pr = C_P\mu/k_g$ is the Prandtl number of the gas phase. $B_T = (1 + B_M)^\varphi - 1$ is the Spalding heat transfer number, in which $\varphi = (C_{p,v}/C_{P,d})/Le$. $Le \equiv \alpha/D$ ($\alpha$ is the thermal diffusivity, whilst $D$ is the mass diffusivity) is the Lewis number of the gaseous mixture and $C_{p,v}$ is the constant pressure specific heat of water vapour. The droplet heating and evaporation models are validated against with the single water droplet experiments (see Appendix A) and good accuracies are demonstrated about the evolutions of droplet diameter and temperature.

The equation of droplet number density $N_d$ reads

$$\frac{\partial N_d}{\partial t} + \frac{\partial (N_d u_d)}{\partial r} + 2\frac{N_d u_d}{r} = 0. \tag{18}$$

In this study, two-way coupling between the gas and droplet phase is considered, characterized by the exchange of species, momentum and energy. They correspond to the individual terms in $S_L$ in Eq. (2)

$$\begin{cases} S_m = N_d \dot{m} \\ S_v = -N_d m_d \frac{u - u_d}{\tau_r} \\ S_e = -N_d h_c A_d (T - T_d) + N_d \dot{m}_d H_g(T_d) \end{cases}. \tag{19}$$

$S_m$ is non-zero only for the equation of H$_2$O mass fraction. $H_g(T_d)$ is the enthalpy of water vapour at droplet temperature.

## 2.3 Physical model

The one-dimensional spherical reactor is shown in Fig. 1. The radius of the domain is $R = 4$ cm. For the gas phase, the initial distributions of pressure, velocity and composition are uniform in the domain. Specifically, the initial pressure $P_0$ is 40 atm, which is close to the actual pressure in a combustion chamber of IC engine [1]. The initial gas velocity is zero, i.e., $u_0 = 0$ m/s. The reactor is filled with stoichiometric *n*-heptane/air mixture.

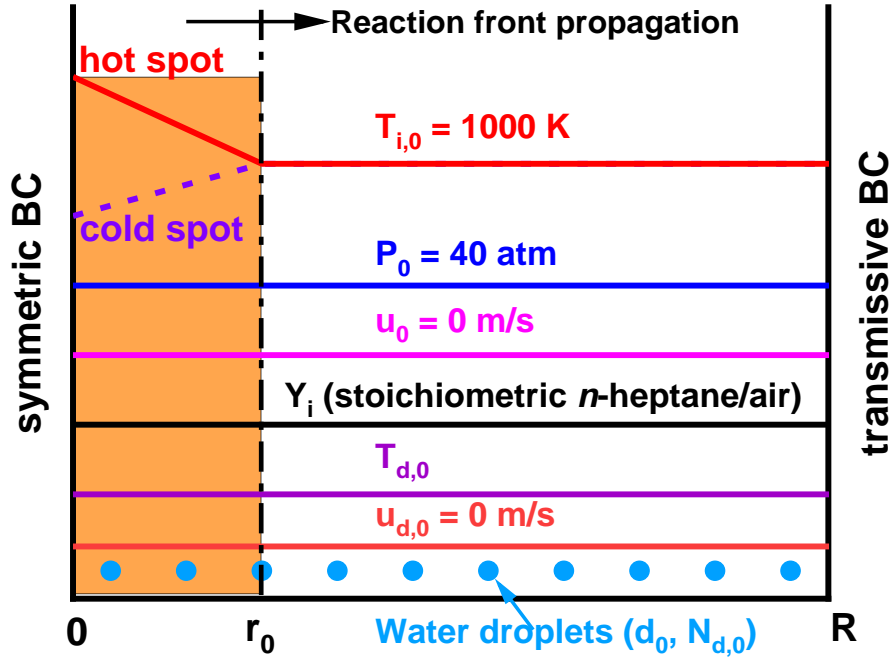

Fig. 1 Schematic of the one-dimensional spherical reactor.

Dispersed ultrafine water droplets are considered to mimic the water mists injected into IC engine cylinder to mitigate the knock intensity and reduce pollutant emissions [1,30–32]. The water droplets are assumed to be mono-sized and the initial droplet diameter is $d_0 = 5$ μm. The initial droplet temperature and velocity are $T_{d,0} = 298$ K and $u_{d,0} = 0$ m/s, respectively. The density and constant pressure specific heat of water droplets are 1,000 kg/m$^3$ and 4,200 J/(kg·K), respectively. Initially they are uniformly distributed in the reactor and the number density $N_{d,0}$ is determined based on the initial droplet diameter $d_0$ and volume fraction $\alpha_{d,0}$, i.e., $N_{d,0} = \alpha_{d,0}/\pi d_0^3/6$. Note that dilute droplet concentrations are studied here and hence the initial volume fraction $\alpha_{d,0}$ in our study is less than 1‰, following Crowe et al. [54].

The reactive front is initiated by an ignition spot with temperature gradient near the left boundary, and therefore the initial temperature $T_0$ in the reactor is

$$T_0(r) = \begin{cases} T_{i,0} + (r - r_0)\frac{dT_0}{dr}, & for\ 0 \leq r \leq r_0 \\ T_{i,0}, & for\ r_0 < r \leq R \end{cases}, \quad (20)$$

where $r_0$ is the radius of the ignition spot, which is fixed to be 3.5 mm in this study. $dT_0/dr$ is the initial temperature gradient inside the spot, which is constant and varied in our simulations. $T_{i,0}$ is the initial gas temperature beyond the spot, which is fixed to be 1,000 K, close to the end gas temperature in IC engines [16,25]. Given that the Negative Temperature Coefficient (NTC) phenomenon may occur due to droplet evaporation cooling, positive or negative initial temperature gradients $dT_0/dr$ are used to initiate the autoigniting wave, which will be discussed in detail in Section 4.

The symmetric condition is enforced at $r = 0$, i.e.,

$$\begin{cases} u = 0, \frac{\partial T}{\partial r} = \frac{\partial Y_i}{\partial r} = 0 \\ u_d = 0, \frac{\partial d}{\partial r} = \frac{\partial T_d}{\partial r} = \frac{\partial N_d}{\partial r} = 0 \end{cases}. \quad (21)$$

At $r = R$, the transmissive condition is used, i.e.,

$$\begin{cases} \frac{\partial u}{\partial r} = \frac{\partial T}{\partial r} = \frac{\partial Y_i}{\partial r} = 0 \\ \frac{\partial u_d}{\partial r} = \frac{\partial d}{\partial r} = \frac{\partial T_d}{\partial r} = \frac{\partial N_d}{\partial r} = 0 \end{cases}. \quad (22)$$

## 2.4 Numerical implementation

The governing equations of gas and liquid phases are solved using a well-validated in-house code A-SURF (Adaptive Simulation of Unsteady Reactive Flow) [55]. This has been proven to be an accurate tool for predicting shock and detonation waves [14,20–24]. The finite volume method is used to discretize the gas phase equations. The Strang splitting fractional-step procedure with second-order accuracy is adopted to separate the time evolution of reaction term $S_R$ from that of the convection term $F(U)$, diffusion term $F_v(U)$ and source/sink term $S_L$ to reduce the overall computational cost. For the liquid phase equations, the first-order accurate explicit Euler scheme is used for temporal discretization. The second-order central differencing scheme is applied for convection terms. Besides, the source terms in Eqs. (3), (11), (14) and (18) are integrated explicitly.

Dynamically Adaptive Mesh Refinement (AMR) algorithm [56] is used to capture the shock / reaction front and the maximum level of refinement is assumed to 9. It is found that further mesh refinement does not have influence of the reaction front evolutions. Besides, the time step is $5\times10^{-11}$ s, which leads to CFL number (based on gas properties) less than 0.4. Moreover, a skeletal $n$-$C_7H_{16}$ mechanism (44 species and 112 reactions) [57] with low-temperature chemistry is used and its capacity in predicting $n$-heptane detonation and low-temperature oxidation has been corroborated in previous studies [20,21,24,37,38,58].

## 3 Chemical explosive mode analysis

The Chemical Explosive Mode Analysis (CEMA) [59–62], inspired by the computational singular perturbation method [63,64], is used to extract the fundamental chemical state in reaction front development process. For a typical chemically reactive flow, the equations can be written as

$$\frac{D\boldsymbol{\varphi}}{Dt} = \boldsymbol{\omega}(\boldsymbol{\varphi}) + \boldsymbol{s}(\boldsymbol{\varphi}), \tag{23}$$

where $D(\cdot)/Dt$ is the material derivative. $\boldsymbol{\varphi}$ is the vector of primary variables consisting of all species and temperature, i.e., $\boldsymbol{\varphi} = [Y_1, \cdots Y_n, T]$. In the RHS of Eq. (23), $\boldsymbol{\omega}(\boldsymbol{\varphi})$ is the vector of the chemical source terms, whilst $s(\varphi)$ is the vector of the non-chemical terms (e.g., diffusion). The CEMA is based on eigen analysis of the Jacobian matrix $J_\omega$ of the chemical source term $\omega(\varphi)$. A chemical mode is defined as an eigenmode of $J_\omega$, which contains an eigenvalue and the corresponding eigenvectors. Furthermore, a CEM mode is the chemical mode whose real part of the eigenvalue $\lambda_e$ is positive, i.e., $\text{Re}(\lambda_e) > 0$. This indicates the propensity of chemical explosion when the mixture is isolated.

The contribution of a chemical species towards a CEM is quantified by the Explosion Index (EI) [61]

$$\mathbf{EI} = \frac{\text{diag}|\boldsymbol{a}_e \boldsymbol{b}_e|}{\text{sum}(\text{diag}|\boldsymbol{a}_e \boldsymbol{b}_e|)}, \tag{24}$$

where $\boldsymbol{a}_e$ and $\boldsymbol{b}_e$ are respectively the right and left eigenvectors, and "diag|·|" denotes the elementwise absolute values. The elements of **EI** range from 0 to 1. Similar EI can also be calculated for temperature. Higher **EI** value indicates higher contribution of the species or temperature in a CEM. Besides, the contribution of a reaction to a CEM is measured by the Participation Index (PI) [61]

$$\mathbf{PI} = \frac{|(\boldsymbol{b}_e \cdot \boldsymbol{S}) \otimes \boldsymbol{R}|}{\text{sum}(|(\boldsymbol{b}_e \cdot \boldsymbol{S}) \otimes \boldsymbol{R}|)}, \tag{25}$$

where $\boldsymbol{S}$ is the stoichiometric coefficient matrix, $\boldsymbol{R}$ is the vector of net reaction rate, "$\otimes$" represents the element-wise multiplication of two vectors. All the elements of **PI** lie within [0,1] and the reaction is dominant in the CEM if its PI is close to unity.

**4 Homogeneous ignitions of two-phase mixtures**

To quantify the autoignition process in an ignition spot with temperature gradient, three parameters are used, i.e. ignition delay time $\tau_{ig}$, excitation time $\tau_e$, and critical temperature gradient $(dT/dr)_c$ [10]. Specifically, $\tau_{ig}$ is the duration when the heat release rate reaches its maximum from the initial instant, whilst $\tau_e$ denotes the time interval from 5% to maximum heat release [10]. Moreover, based on the theories by Zel'dovich [8] and Gu et al. [10], $(dT/dr)_c$ quantifies a critical temperature gradient within the ignition spot for chemical resonance and hence detonation development, i.e.,

$$\left(\frac{dT}{dr}\right)_c = \left[a\left(\frac{d\tau_{ig}}{dT_0}\right)\right]^{-1}, \tag{26}$$

where $a = \sqrt{kR_gT}$ is the sound speed. $k$ is the adiabatic index and $R_g$ is the gas constant. Under this temperature gradient, the reaction front from the ignition spot propagates at the speed of sound.

Homogeneous ignitions of stoichiometric $n$-C$_7$H$_{16}$/air mixtures with ultrafine water droplets in the 1D reactor in Fig. 1 will be discussed in Sections 4.1−4.3, to evaluate the effects of the droplet diameter and volume fraction on the abovementioned parameters, i.e., $\tau_{ig}$, $\tau_e$, and $(dT/dr)_c$. Here all initial variables of gas and liquid phases are spatially uniform, thereby leading to zero-dimensional simulations in nature.

## 4.1 Homogeneous ignition process with droplets

Figure 2 shows the ignition delay time and critical temperature gradient of droplet-laden $n$-C$_7$H$_{16}$/air mixture versus gas temperature. The initial droplet volume fraction is $\alpha_{d,0} = 8.0\times10^{-4}$. The results from the droplet-free stoichiometric $n$-C$_7$H$_{16}$/air mixture are also included. It is seen that the droplet addition considerably affects the dependence of $\tau_{ig}$ and $(dT/dr)_c$ on gas temperature. The turnover temperature (marked as symbols on Fig. 2a) [65] is increased significantly in the two-phase mixtures compared to those of the droplet-free mixtures. Specifically, the lower turnover temperature is $T_l = 1,000$ K, whereas the upper turnover temperature is $T_u$ = 1,100 K, larger than the counterparts (850 and 950 K) of droplet-free mixtures. Because of the NTC effects, the distributions of $(dT/dr)_c$ have three sections: middle positive branch and two lower negative ones, as shown in Fig. 2(b). A negative (positive) $(dT/dr)_c$ indicates that a hot (cold) spot is required for detonation development in 1D simulations [23].

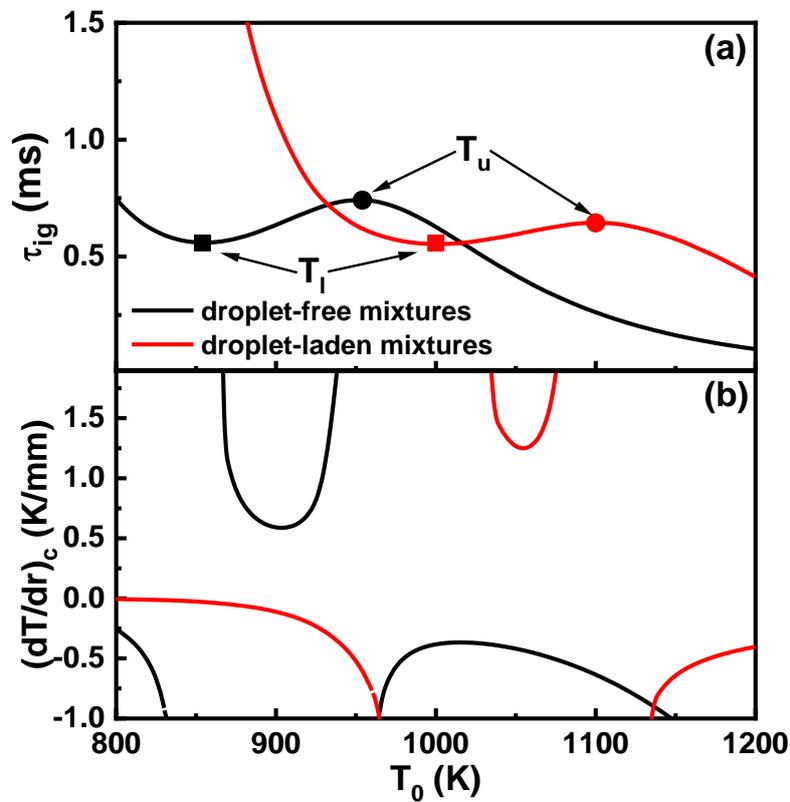

Fig. 2 (a) Ignition delay and (b) critical temperature gradient of stoichiometric $n$-$C_7H_{16}$/air mixtures with water droplets. $\alpha_{d,0}$ = 8.0×10$^{-4}$. $T_u$ and $T_l$: upper and lower turnover temperature.

Figure 3 shows the time histories of temperature, pressure, heat release rate, and droplet diameter in homogeneous ignition of stoichiometric $n$-$C_7H_{16}$/air mixtures with water droplets. The initial volume fraction is 8.0×10$^{-4}$, and the initial temperature is 1,000 K. One can see that, due to the heat absorption by water droplets, the gas temperature is reduced to a minimum value of 840.6 K at $t$ = 0.308 ms, and the low-temperature chemistry is therefore initiated.

Three ignition events are observed from the corresponding heat release peaks, respectively at $t$ = 0.465, 0.536 and 0.554 ms. They are respectively termed as low-, intermediate- and high-temperature ignitions (abbreviated as LTI, ITI, and HTI) [21]. It is noteworthy that the water droplets complete the evaporation at $t$ = 0.308 ms, which is ahead of the onset of the LTI. Therefore, all the three ignition events occur in purely gaseous environment, although the initial mixture is laden with the ultrafine water droplets.

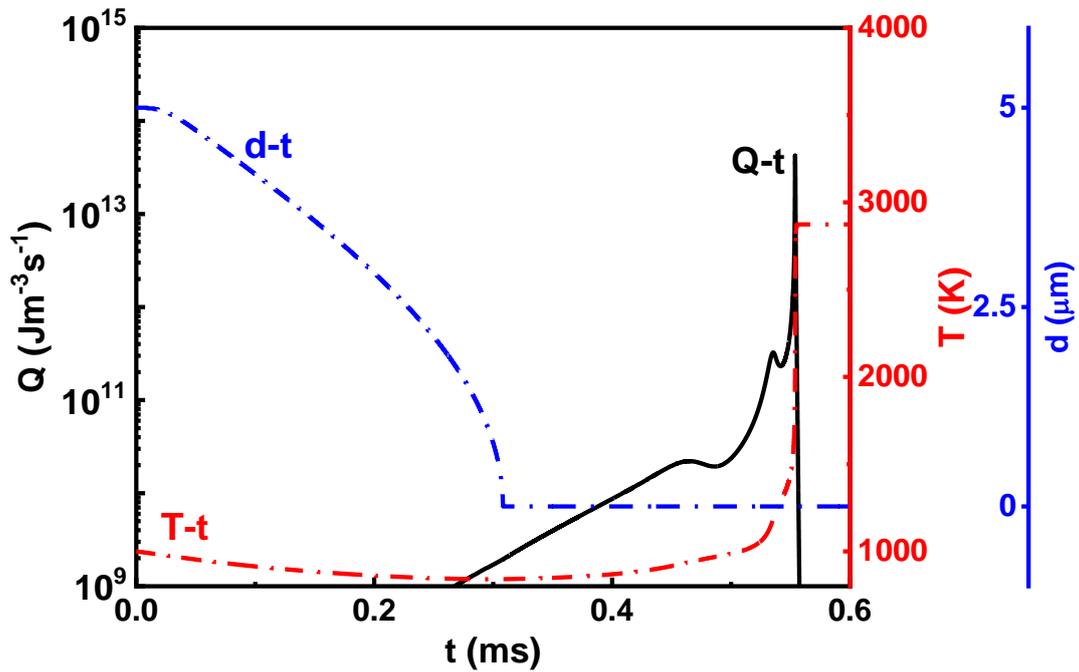

Fig. 3 Time history of pressure, temperature, heat release rate, $H_2O$ mass fraction and droplet diameter of stoichiometric $n$-$C_7H_{16}$/air mixture with droplets. $\alpha_{d,0}$ = 8.0×10$^{-4}$ and $T_0$ = 1,000 K.

In the above discussion, the droplet volume fraction is fixed to be 8.0×10$^{-4}$ and its effects on $n$-$C_7H_{16}$/air autoignition will be further studied in Fig. 4, which shows the ignition delay time, droplet evaporation time and turnover temperature with volume fractions of 0.005%−0.1%. For multi-stage autoignition, the ignition delay time of each stage is defined as the instant of maximum heat release with respect to the initial time, as shown in Fig. 3. $\tau_{ig,1}$, $\tau_{ig,2}$ and $\tau_{ig,3}$ (i.e., $\tau_{ig}$ in Fig. 2) are the corresponding ignition delay times for LTI, ITI and HTI, respectively. It is seen from Fig. 4(a) that only high-temperature ignition is observed when $\alpha_{d,0} < 5.0 \times 10^{-4}$. With $\alpha_{d,0} \geq 5.0 \times 10^{-4}$, multi-stage ignition appears. Increased $\alpha_{d,0}$ leads to monotonically increased $\tau_{ig,1}$. However, $\tau_{ig,2}$ and $\tau_{ig,3}$ show non-monotonic change with $\alpha_{d,0}$, indicating the NTC behaviours due to the water droplet evaporation. Moreover, the droplet evaporation time $\tau_{evap}$ is also shown, which corresponds to the instant when the droplet diameters in the reactor are reduced to $d \leq 10^{-12}$ μm (hence deemed complete evaporation). It is coloured by the gas phase temperature. It is found that $\tau_{evap}$ is much smaller than the LTI ignition delay for all the shown range of $\alpha_{d,0}$. This indicates that ignition proceeds in purely gaseous mixtures. The gas temperature when the droplets are fully vaporized in the reactor is also marked in the curve of $\tau_{evap}$ and it ranges from 980 K to 800 K when $\alpha_{d,0} < 1.0 \times 10^{-3}$. Moreover, it is seen from Fig. 4(b) that both lower and upper turnover temperatures increase with $\alpha_{d,0}$, implying that the NTC region moves towards higher temperature with increased $\alpha_{d,0}$. At about $\alpha_{d,0} = 8.0 \times 10^{-4}$, the value of $T_l$ is close to the initial temperature of the gas mixture $T_0 = 1000$ K.

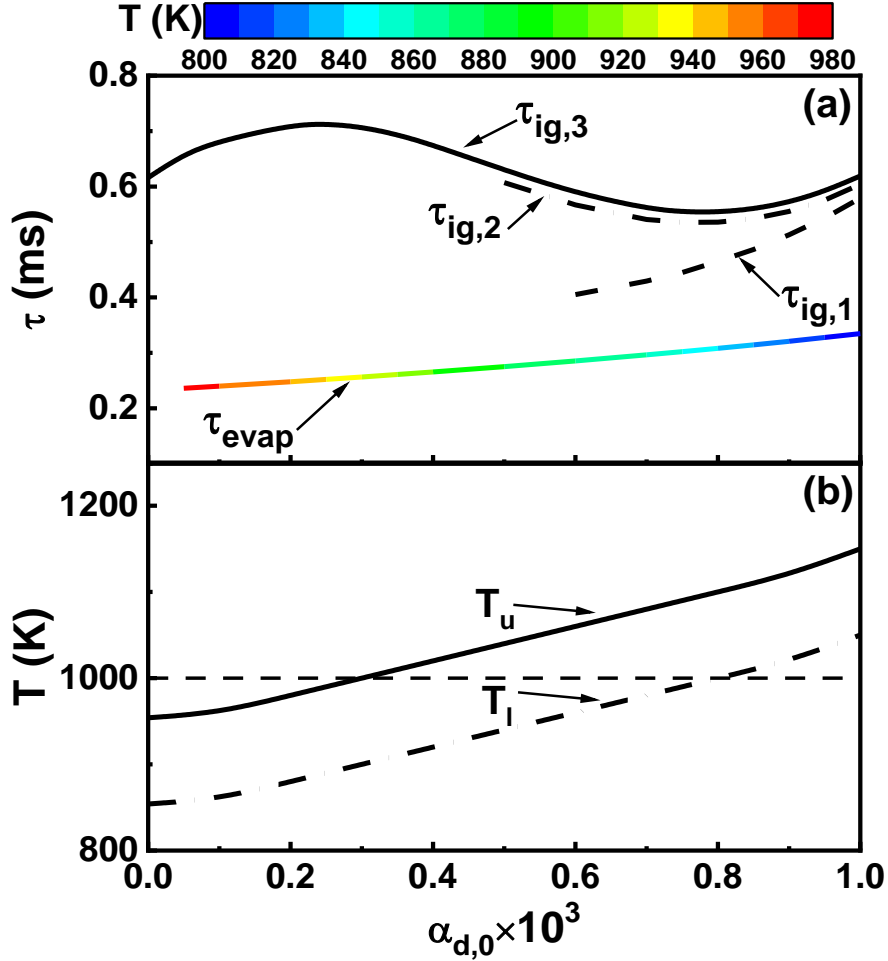

Fig. 4 (a) Ignition delay time and droplet evaporation time and (b) turnover temperature as functions of initial volume fraction. $T_l$ is the lower turnover temperature, $T_u$ is the upper turnover temperature. $T_0 = 1{,}000$ K.

Figure 5 shows the excitation time $\tau_e$ and critical temperature gradient $(dT/dr)_c$ as functions of initial droplet volume fraction $\alpha_{d,0}$. To evaluate the water vapour dilution effects, fully pre-vaporized results (fully vaporized droplets in stoichiometric $n$-$C_7H_{16}$/air mixture with temperature of 1,000 K. Water steam mass fractions $Y_{H2O}$ marked at the top $x$-axis) are also added. It is seen from Fig. 5(a) that the excitation time $\tau_e$ increases monotonically with $\alpha_{d,0}$. This is because the increased effects of cooling and dilution with more droplet evaporation. The difference of $\tau_e$ between droplet-laden and fully pre-vaporized mixtures is caused by the cooling effect. The relative errors of excitation time, corresponding to some selected $\alpha_{d,0}$, induced by cooling effect are marked in Fig. 5(a). One can see that the cooling effect becomes more crucial with increased $\alpha_{d,0}$.

Figure 5(b) shows the critical temperature gradient as a function of initial volume fraction. It is noted that the critical temperature gradient is associated with ignition delay time (see Eq. 26). Thus, for multi-stage autoignition, the critical temperature gradient of different stages can be obtained. It is seen from Fig. 5(b) that three branches of the critical temperature gradient of HTI exist, and a negative (positive) value of $(dT/dr)_c$ indicates that a hot (cold) spot is required for simulations of inhomogeneous mixture in Section 5. Besides, two branches of the critical temperature gradient of ITI are observed, whilst only one branch exists for the critical temperature gradient of LTI.

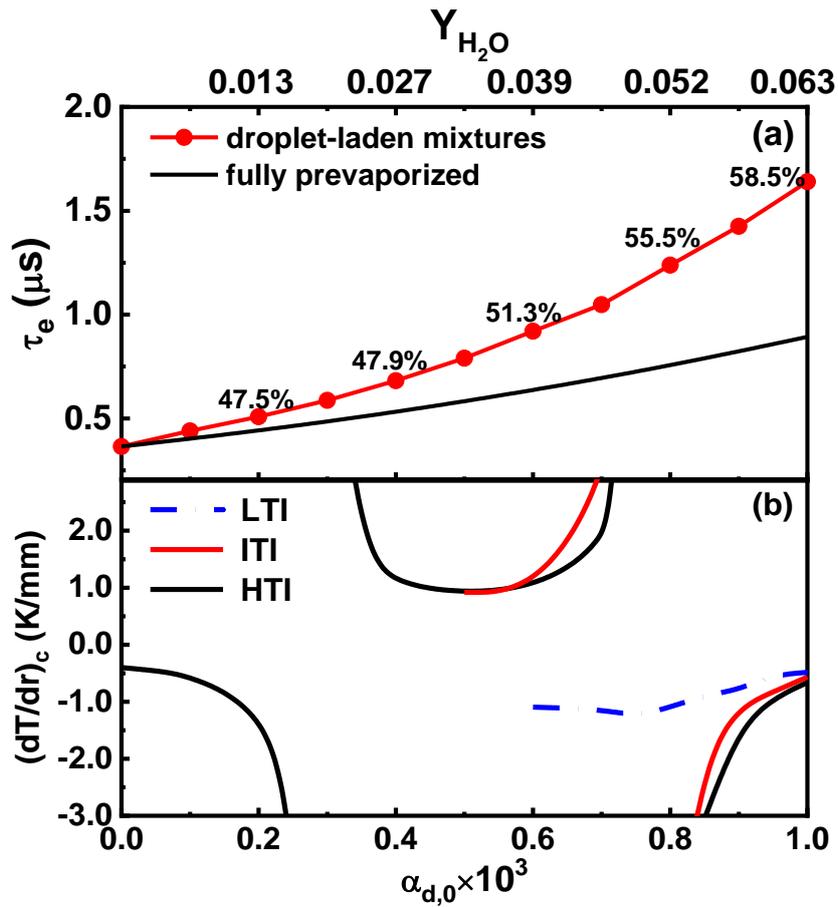

Fig. 5 (a) Excitation time and (b) critical temperature gradient as functions of initial volume fraction. $T_0 = 1,000$ K. $Y_{H_2O}$ along the top axis indicates the water vapour mass fraction when the droplets are fully vaporized.

**4.2 Chemical explosive mode analysis**

To further understand the multi-stage autoignition of two-phase $n$-$C_7H_{16}$/air mixture, Fig. 6 shows the time evolutions of the real part of the eigen value of chemical Jacobian matrix, $\text{Re}(\lambda_e)$, and EIs of the dominant species. It corresponds to the results in Fig. 3. It is seen that zero-crossing of $\text{Re}(\lambda_e)$ is observed for the LTI and HTI events. This is also observed in Ref. [60,66] for $n$-heptane autoignition. Moreover, as seen from Fig. 6(b), the contributions of $HO_2$ or $OC_7OOH$ (KET) towards CEM are dominant prior to the LTI. Nevertheless, when LTI is initiated, $C_2H_2$ becomes most important. This indicates that the LTI is mainly controlled by radical proliferation [59–61]. After LTI, the temperature contribution becomes dominant (see red circles in Fig. 6), corresponding to the thermal runaway process [59–61].

To reveal the contributions of the individual elementary reactions towards the CEM, Fig. 6(c) shows the time evolutions of the PI's of dominant elementary reactions. The related chemical reactions are listed in Table 1. It is seen from Fig. 6(c) that, before LTI (marked as a solid square along the time axis), R104, R106 and R107 (see reactions in Table 1) are dominant, which correspond to $n$-$C_7H_{16}$ oxidation and generation of R, $RO_2$ and QOOH, indicating that the low-temperature chemistry is crucial during this period. After LTI, $C_0$ (R7 and R15) and $C_2$ (R60) oxidation become dominated during ITI and HTI. As such, the results in Fig. 6(c) further confirm the multi-stage ignition induced by the evaporation of dispersed ultrafine water droplets with intermediate or high loadings (see Fig. 4).

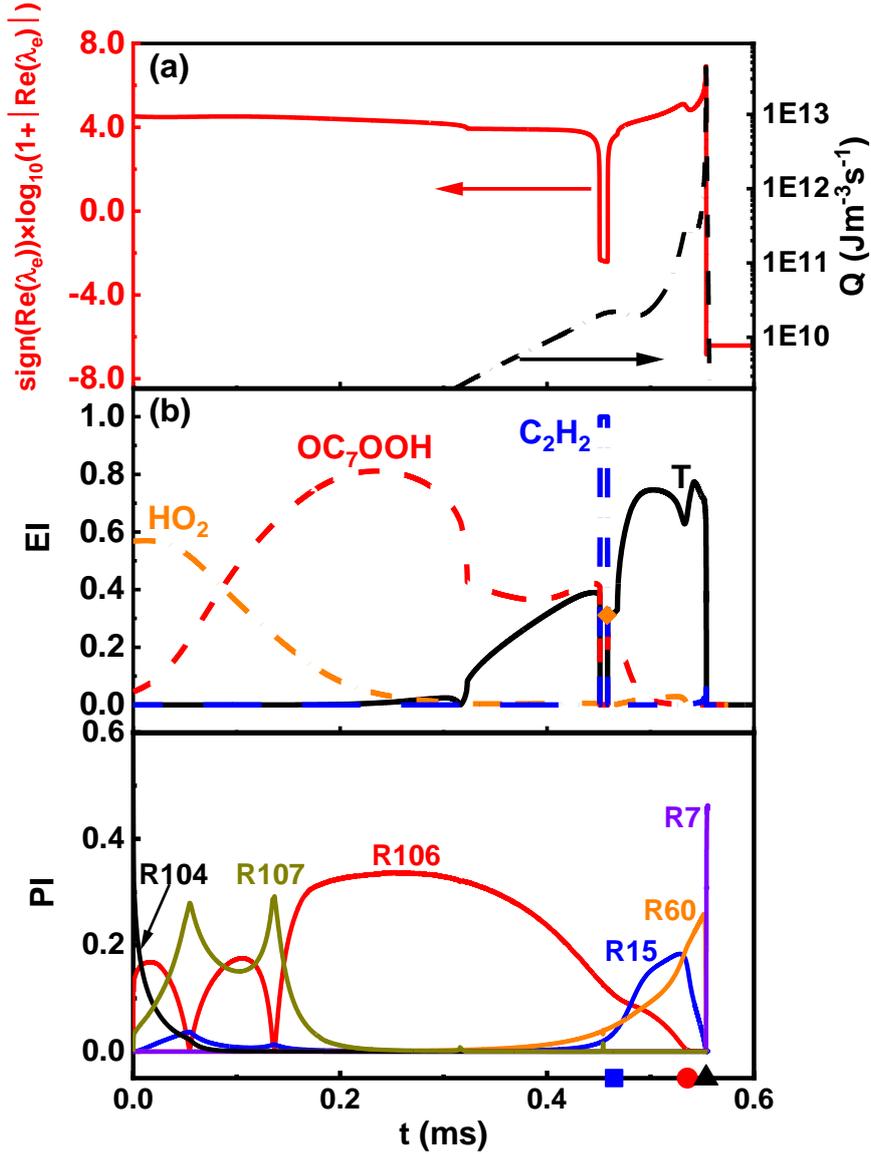

Fig. 6 Time evolutions of (a) real part of eigenvalue $\text{Re}(\lambda_e)$, heat release rate, (b) EI's, and (c) PI's of dominant elementary reactions. $\alpha_{d,0} = 8.0 \times 10^{-4}$ and $T_0 = 1,000$ K. Orange diamond in (b): onset of thermal runaway process. Symbols on $x$-axis: blue square: $\tau_{ig,1}$; red circle: $\tau_{ig,2}$; black triangle: $\tau_{ig,3}$.

Table 1 Dominant elementary reactions identified with CEMA

| Index | Reaction |
|---|---|
| R7 | $H+OH+M\Leftrightarrow H_2O+M$ |
| R15 | $2OH+M\Leftrightarrow H_2O_2+M$ |
| R60 | $C_2H_3+O_2 \Rightarrow CH_2O+HCO$ |
| R104 | $NXC_7H_{16}+O_2 \Rightarrow SXC_7H_{15}+HO_2$ $(RH+O_2 \Rightarrow R+HO_2)$ |
| R106 | $SXC_7H_{15}+O_2 \Leftrightarrow PC_7H_{15}O_2$ $(R+O_2 \Leftrightarrow RO_2)$ |
| R107 | $PC_7H_{15}O_2 \Rightarrow PHEOOHX_2$ $(RO_2 \Rightarrow QOOH)$ |

# 5 Reaction front development from an ignition spot in two-phase mixtures

## 5.1 Reaction front initiation mode

Figure 7 summarizes the reaction front initiation modes from our 1D numerical simulations with an ignition spot. They are identified based on the relations between gas temperature $T_{i,0}$ and lower turnover temperature $T_l$. It is seen from Fig. 2(a) that the corresponding lower turnover temperature, $T_l$, is 1,000 K at about $\alpha_{d,0} = 8.0 \times 10^{-4}$, which equals the initial gas temperature outside the ignition spot, $T_{i,0}$. Besides, the effect of droplet volume fraction on turnover temperature is also shown in Fig. 4(b). Below are the descriptions for the various modes:

1) If $T_{i,0}$ is close to $T_l$ (see Fig. 7a), then the ignition delay time at the right end of the ignition spot, $\tau_{ig,3}(T_{i,0})$, is always shorter than that at the left, $\tau_{ig,3}(T_0(r=0))$, regardless of negative or positive temperature gradient inside the spot. Therefore, autoignition is initiated at the right of the ignition spot, and the HTI waves travel from right to left within the ignition spot, leading to an implosion over the ignition spot. This is termed as mode a.

2) If $T_{i,0}$ is slightly higher (lower) than $T_l$ (Figs. 7b and 7c), then a cold (hot) spot is required to initiate the reaction front. One can see that the lower turnover temperature $T_l$ is reached at some locations inside the ignition spot. Thus, the HTI waves are initiated inside the ignition spot and subsequently two oppositely propagating HTI waves are formed. Figures 7(b) and 7(c) correspond to modes b and c.

3) If the initial end gas temperature $T_{i,0}$ is sufficiently higher (lower) than $T_l$ (Figs. 7d and 7e), then the gas temperature reduced by droplet evaporation is beyond the NTC temperature range. As such, the HTI waves are initiated at the left boundary of the ignition spot and propagate rightward. They are modes d and e.

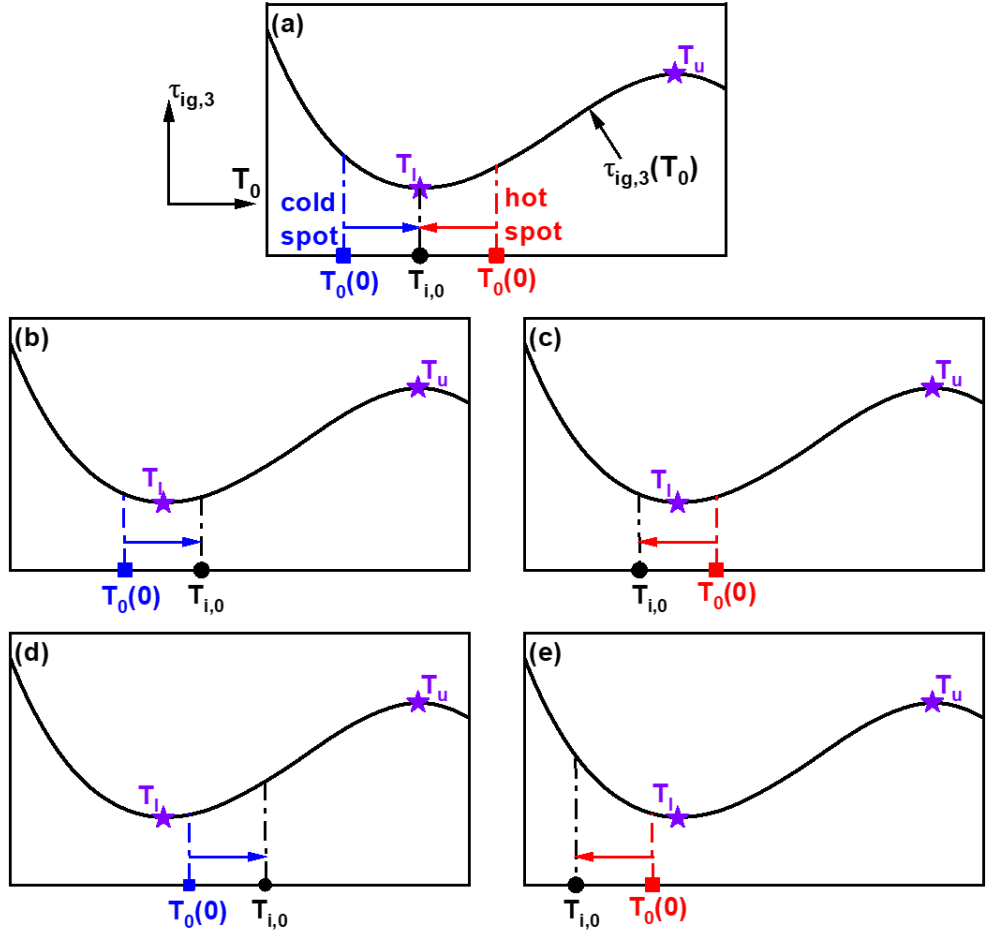

Fig. 7 Reaction front initiation mode: (a) $T_{i,0}$ is close to $T_l$ (i.e., $T_0(0) < T_{i,0} \approx T_l$ or $T_{i,0} \approx T_l < T_0(0)$), (b) $T_{i,0}$ is slightly higher than $T_l$ ($T_0(0) < T_l < T_{i,0}$), (c) $T_{i,0}$ is slightly lower than $T_l$ ($T_{i,0} < T_l < T_0(0)$), (d) $T_{i,0}$ is sufficiently higher than $T_l$ ($T_l < T_0(0) < T_{i,0}$), and (e) $T_{i,0}$ is sufficiently lower than $T_l$ ($T_{i,0} < T_0(0) < T_l$). $T_{i,0}$ is initial gas temperature outside the ignition spot, whereas $T_0(0)$ is the initial gas temperature at the left boundary.

Figure 8 further characterizes the reaction front initiation mode in terms of reaction front propagating direction and speed, under various temperature gradients and droplet volume fractions. The temperature gradient is normalized based on the maximum one of the corresponding volume fractions, i.e., $(dT_0/dr)_{\alpha_{d,0},\max}$. For a constant $\alpha_{d,0}$, the $(dT_0/dr)_{\alpha_{d,0},\max}$ is selected, such that it is large enough to cover all the possible reaction front initiation modes. For instance, they are 25.71, -11.79, and -9.98 K/mm for $\alpha_{d,0} = 8.0\times10^{-4}$, $9.0\times10^{-4}$, and $1.0\times10^{-3}$, respectively. It is seen that the autoignition waves initiated at the left end of the ignition spot (i.e., mode d and e in Figs. 7d and 7e) are found for most occasions (square and diamond symbols). Furthermore, when the temperature gradient is increased, the propagation speed of the autoignition wave is changed from supersonic to subsonic conditions for a constant droplet loading, e.g., $\alpha_{d,0} = 2.0\times10^{-4}$. This is because the interaction between pressure wave generated by heat release rate and autoignition wave is weakened. Besides, the autoignition wave from the right end of the ignition spot (i.e., mode a in Fig. 7a) is found for $\alpha_{d,0} = 8.0\times10^{-4}$, because the corresponding initial gas temperature outside the ignition spot is almost equal to the lower turnover temperature. Furthermore, change of the reaction front initiation mode is also found for $\alpha_{d,0} = 7.0\times10^{-4}$ and $9.0\times10^{-4}$. With a higher initial temperature gradient, autoignition waves traveling towards both left and right sides (i.e., mode b and c in Figs. 7b and 7c) are observed. This is because the lower turnover temperature $T_l$ is reached at some locations inside the ignition spot as described in Fig. 7(b) and (c). Whether the results from the two-phase gas-droplets can be accommodated in the Bradley's diagram [9,10] will be discussed in Section 5.5.

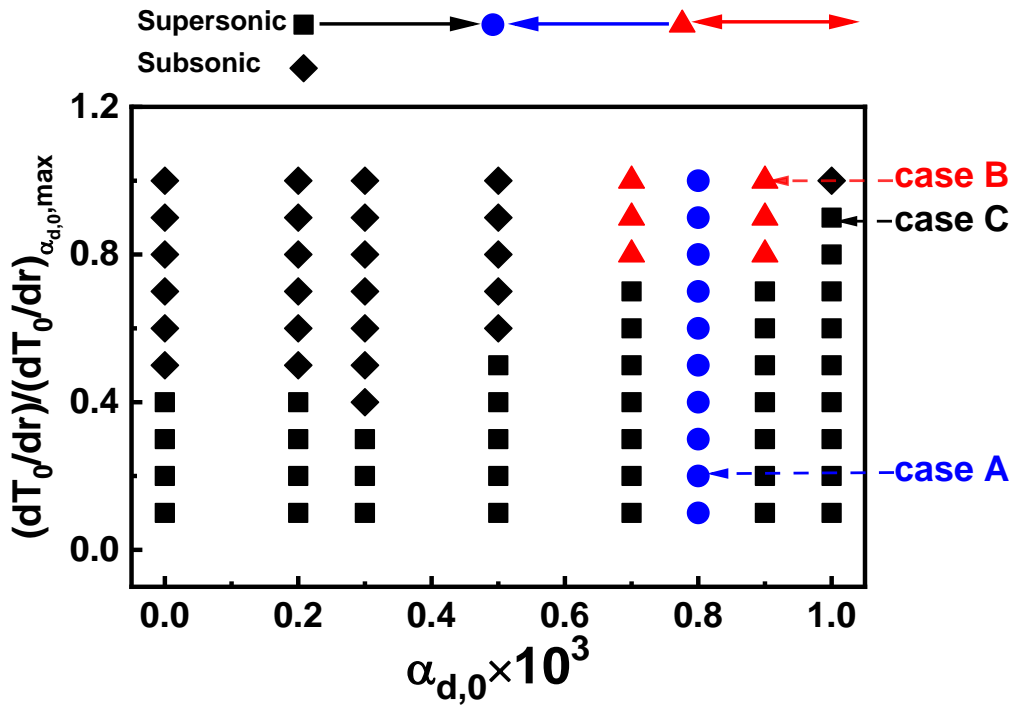

Fig. 8 Reaction front initiation mode in the diagram of ignition spot temperature gradient versus droplet volume fraction. $d_0 = 5$ μm and $T_0 = 1,000$ K. Ignition spot size is $r_0 = 3.5$ mm. Black diamonds: subsonic wave from left to right, Black squares: supersonic wave from left to right, Blue circles: from right to left, and red triangles: for middle to both sides.

## 5.2 Reaction front development within the ignition spot

Reaction front development within an ignition spot in two-phase medium will be studied in this section, through three representative cases, i.e., A, B and C, tabulated in Table 2 and marked in Fig. 8. Their spot radii are $r_0$ = 3.5 mm, and their droplet volume fractions are 8.0×10$^{-4}$, 9.0×10$^{-4}$ and 1.0×10$^{-3}$, respectively. Specifically, a cold spot (i.e., $dT_0/dr > 0$) is needed for case A, whilst a hot spot (i.e., $dT_0/dr < 0$) for cases B and C. The corresponding temperature gradient and reaction front initiation mode of cases A−C are listed in Table 2.

Table 2 Information of cases A−C

| Case | $\alpha_d$ | $dT_0/dr$ (K/mm) | Reaction front initiation mode |
| --- | --- | --- | --- |
| A | 8.0×10$^{-4}$ | 5.14 | a |
| B | 9.0×10$^{-4}$ | -11.79 | c |
| C | 1.0×10$^{-3}$ | -7.98 | e |

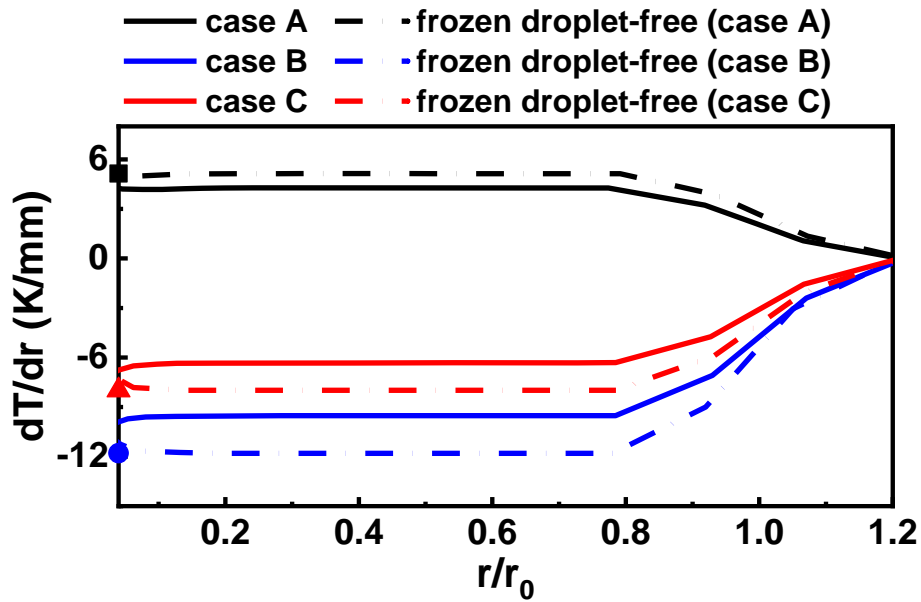

Fig. 9 Spatial distributions of gas temperature gradient when the droplets are critically vaporized. The corresponding initial gas temperature gradients of cases A−C are marked as symbols on the *y* axis.

To reveal the effects of droplet evaporation on the change of temperature gradient within the ignition spot ($r/r_0 \leq 1$) before ignition, Fig. 9 shows the spatial distributions of the gas temperature gradients in cases A−C when all droplets are critically vaporized in the reactor. Apparently, the distributions of temperature gradient at $r/r_0 < 0.8$ are nearly uniform (consistently varies), and the corresponding values are almost 80% of their respective initial temperature gradients (marked as symbols along the *y*-axis). It is noted that the gas temperature gradients may be affected by convection, diffusion, and droplet evaporation (chemical heat release is still weak at this stage). To find out the dominant factor, frozen droplet-free mixtures are considered. We conducted nonreacting droplet-free numerical experiments with same gas mixtures and temperature gradients of cases A−C. Thus, the gas temperature gradients can only be affected by convection and diffusion in these cases. Their counterpart results are also shown in Fig. 9. For the chemically frozen droplet-free mixtures, the corresponding values of gas temperature gradient at $r/r_0 < 0.8$ is almost the same as the respective initial values. Therefore, one can confirm that reduction of the gas temperature gradient at $r/r_0 < 0.8$ is mainly caused by the droplet evaporation cooling effects. Besides, the distributions of temperature gradient at $r/r_0 \geq 0.8$ are mainly controlled by the effects of diffusion, because the differences of temperature near the vicinity of the ignition spot.

Figures 10−12 show the temporal evolutions of temperature and mass fractions of key species within the ignition spot ($r < r_0$) for cases A−C. The zero-crossing points (marked as symbols) of the eigenvalue Re($\lambda_e$) denote the reaction fronts [59–62]. Line #1 corresponds to the early stage of autoignition when the whole droplets in the 1D reactor are critically vaporized. The corresponding temperature gradients are shown in Fig. 9. Lines #2 and #3 correspond to the instants of LTI and ITI at the midpoint of the ignition spot, whilst the rest visualize the HTI process. In Fig. 10, the gas temperature at early stage (line #1) and LTI (line #2) is below 1,000 K due to the heat absorption by the evaporating droplets. It is noted that two zero-crossing

instants for the eigenvalue $\text{Re}(\lambda_e)$ are found at LTI, respectively located at $r \approx 0.11 r_0$ and $0.24 r_0$ of line #2, which are associated with NTC. This is also seen in the 0D results in Fig. 6(a). At ITI (line #3), the maximum gas temperature within the ignition spot is about 1,200 K. For the developments of HTI, autoignition occurs near the right end of the ignition spot (line #4). Subsequently, an autoignition wave travels from right to left within the ignition spot, i.e., mode a as indicated in Fig. 7.

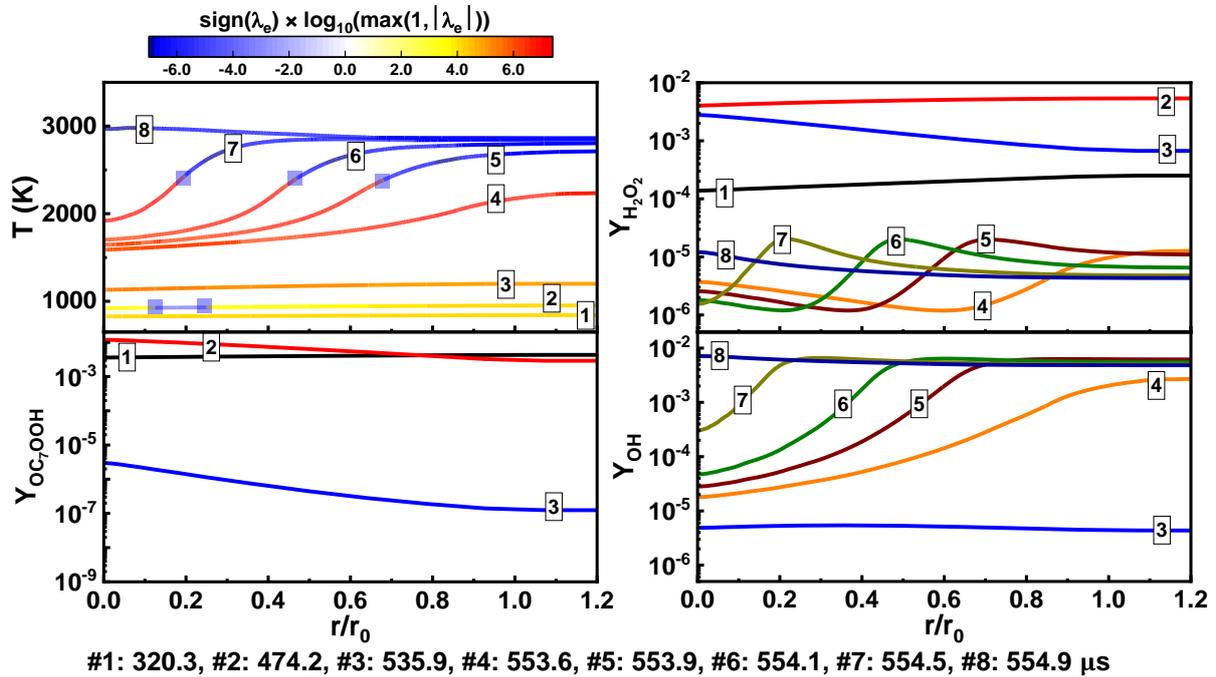

#1: 320.3, #2: 474.2, #3: 535.9, #4: 553.6, #5: 553.9, #6: 554.1, #7: 554.5, #8: 554.9 μs

Fig. 10 Temporal evolutions of temperature and key species mass fractions within the ignition spot in case A. Symbols: zero-crossings of the eigenvalue $\text{Re}(\lambda_e)$.

The NTC phenomenon initiated by the droplet evaporation cooling can also be confirmed through the evolutions of key species. For instance, the mass fractions of $H_2O_2$ and $OC_7OOH$ reach their peaks around LTI (line #2) in the ignition spot, indicating that low-temperature chemistry proceeds during this period. Besides, OH radical is accumulated since ITI (line #3). For HTI process, the peak value of $H_2O_2$ mass fraction is two orders of magnitude less than that at LTI. The OH mass fraction reaches its peak and evolves as the HTI wave propagates outwardly.

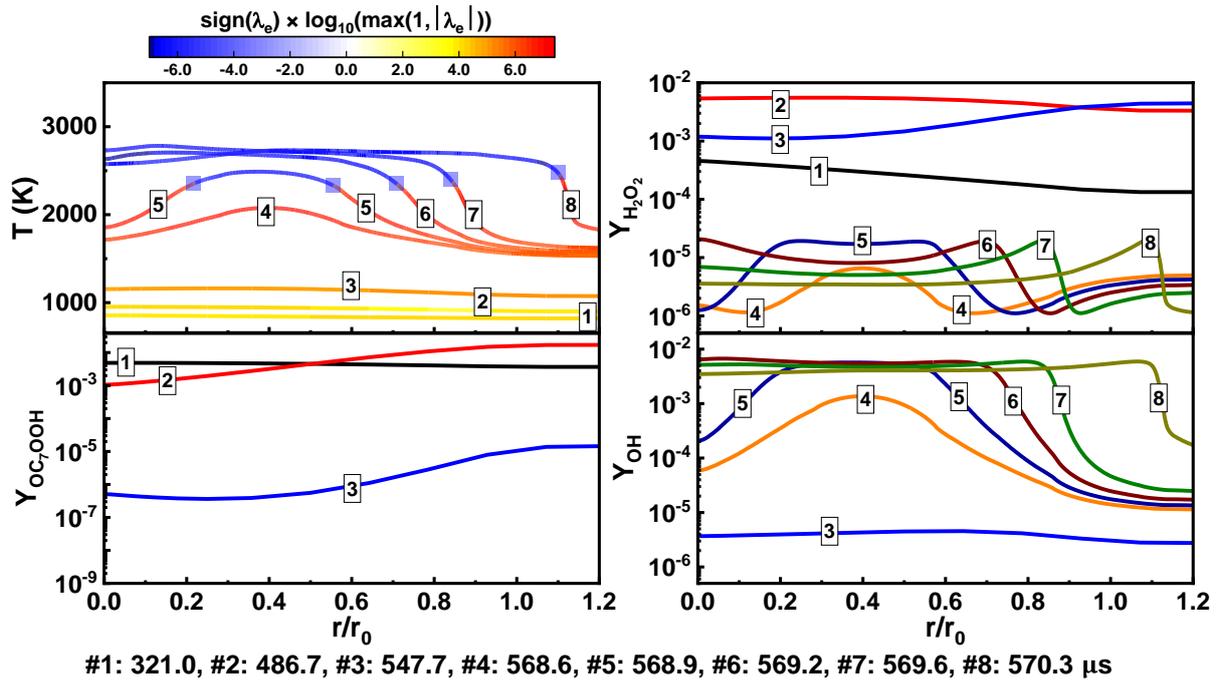

#1: 321.0, #2: 486.7, #3: 547.7, #4: 568.6, #5: 568.9, #6: 569.2, #7: 569.6, #8: 570.3 μs

Fig. 11 Temporal evolutions of temperature and key species mass fractions within the ignition spot in case B. Symbol legend same as in Fig. 10.

Figure 11 shows the counterpart results from case B, which corresponds to mode c. Compared with case A, similar profiles of temperature are found before HTI (lines #1-3). At ITI (line #3), the maximum gas temperature within the ignition spot is about 1,162 K. However, the HTI occurs inside the ignition spot located at $r/r_0 \approx 0.4$ (line #4). Subsequently, two autoignition waves are generated and propagate oppositely (line #5). Note that the temperature of the left-propagating autoignition wave is slightly higher due to the wall compression effects (line #5). Finally, the left-propagating autoignition wave disappears when the reactive gas is fully consumed near the left wall. In terms of the key radicals, corresponding profiles before HTI (lines #1-3) are found to be similar with those of case A. For the HTI development, the OH mass fraction reaches its peak inside the ignition spot and evolves both leftward and rightward when the HTI waves propagate.

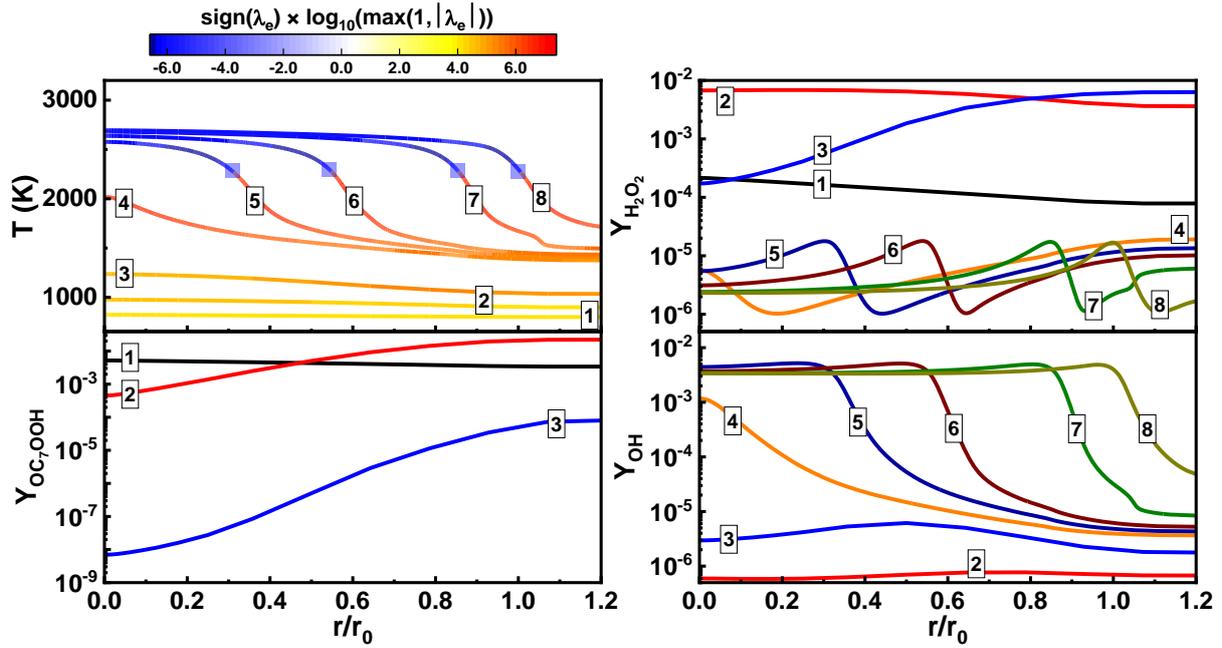

**#1: 335.2, #2: 553.9, #3: 592.2, #4: 610.7, #5: 611.9, #6: 612.9, #7: 614.3, #8: 615.2 μs**

Fig. 12 Temporal evolutions of temperature and key species mass fractions within the ignition spot in case C. Symbol legend same as in Fig. 10.

Moreover, Fig. 12 shows the counterpart results from case C (Mode e). Before HTI (lines #1-3), the evolutions of temperature and radicals are also similar with those of cases A and B. The maximum gas temperature within the ignition spot at ITI (lines 3) is about 1,236 K. It is seen that the HTI occurs firstly at the left end (line #4). Accordingly, the right-ward propagating autoignition wave is observed (lines #5-8), which can be confirmed by the histories of both temperature and OH mass fraction.

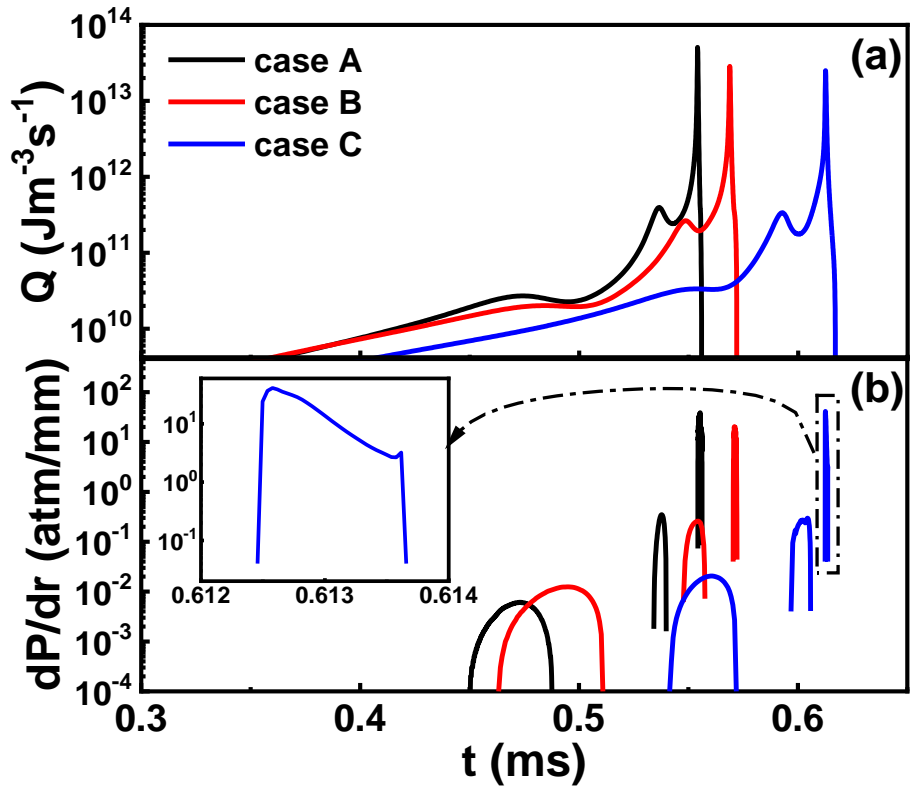

Fig. 13 Time history of (a) heat release rate and (b) pressure gradient of cases A to C.

One can see from Figs. 10−12 that multi-stage ignition occurs in the foregoing three initiation modes. Figure 13 shows the time history of heat release rate and pressure gradient from a probe at the middle of the ignition spot (i.e., $r = r_0/2$) in cases A−C. The pressure gradient history is shown to visualize the pressure wave development. Three pressure waves are observed from Fig. 13(b), corresponding to the three-stage ignition process shown in Fig. 13(a). They can be termed as LTI, ITI, and HTI pressure waves from left to right, respectively. We can see that the magnitudes of LTI pressure wave and ITI pressure wave are at least two orders lower than that of HTI pressure wave. Thus, the HTI pressure wave is more important, which is a key factor for autoignition and detonation development.

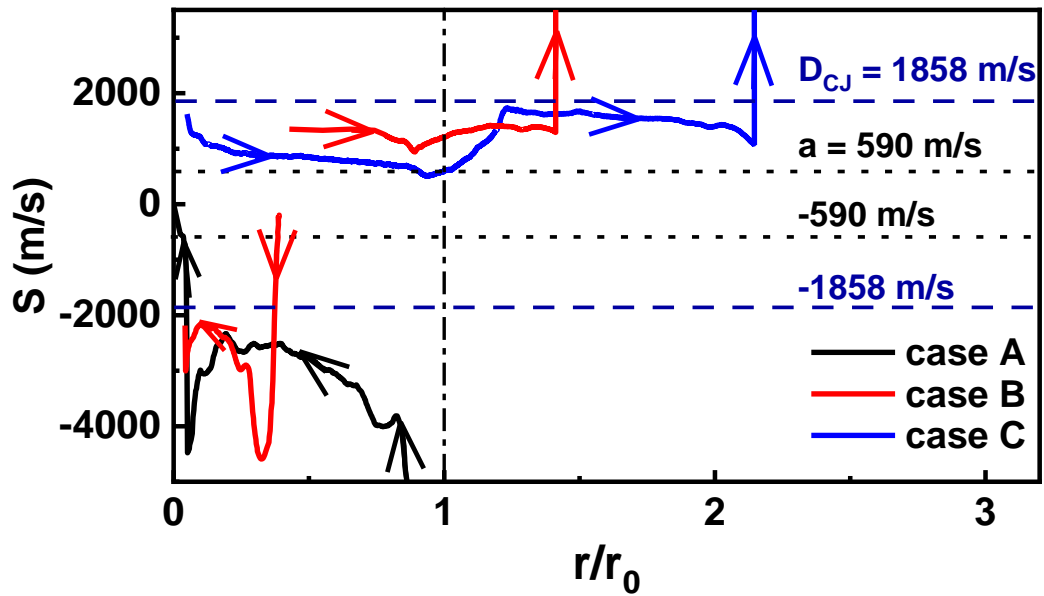

Fig. 14 Reaction front propagation speed as a function of radial coordinate in cases A − C. $D_{CJ}$ is the C−J speed and $a$ is the sound speed.

### 5.3 Reaction front propagation

Figure 14 shows the reaction front propagation speed in cases A−C. The reaction front is extracted from the location with maximum heat release rate. One can see that, in case A, the autoignition wave induced by thermal explosion near the right end of the spot ($r/r_0 = 1$) has a very high initial speed. It travels supersonically from right to left within the ignition spot. Near the left wall, the speed first increases and then decreases quickly to zero. The increase is caused by the wall compression effects, whilst the reduction at $r/r_0 < 0.053$ is because the reactive gas is gradually consumed near the left boundary.

In case B, autoignition wave is initiated at $r/r_0 \approx 0.39$. Two opposite autoignition waves are formed. On one hand, the rightward propagating wave speed is supersonic, but lower than the C−J speed. The reaction front accelerates abruptly when the autoignition of mixture near the right boundary occurs (i.e., $r/r_0 \approx 1.4$ in case B). On the other hand, the leftward propagating wave is faster, and the average speed is about 3,000 m/s, and finally decays near the left wall. The latter is faster because of the higher local heat release caused by the compression effect.

In case C, the reaction front within the ignition spot propagates supersonically and the average propagation speed is about 800 m/s. The reaction front accelerates to the C−J speed outside the ignition spot. The predicted wave speed in case C is approximately 1,525 m/s. It is lower than the C−J speed of stoichiometric droplet-free $n$-$C_7H_{16}$/air mixtures. That may be because of the curvature effects from the spherical geometry [67,68], partial reaction of the end gas before the arrival of the detonation wave, and/or the dilution (mass transfer) and cooling (heat transfer) induced by water droplets evaporation. Note that interphase momentum exchange is not possible since the droplets have been fully gasified before autoignition. Finally, the reaction front accelerates abruptly when the autoignition of mixture near the right boundary occurs (i.e., $r/r_0 \approx 2.1$ in case C).

**5.4 Thermal state**

To analyse the interactions between chemical reaction and pressure waves, Fig. 15 shows the evolutions of thermal states of cases A to C, which are extracted with the aid of a Lagrangian particle initially at the midpoint of the ignition spot. The position of the particle is updated in each time step based on the local flow speed. Therefore, the instantaneous thermal states (e.g., pressure, density, and heat release rate) at the particle position can be obtained from linear extrapolation of the gas properties [21,69]. Here points a, b and c in Fig. 15 denote three ignition stages of LTI, ITI and HTI, respectively. In case A, within the ignition spot, one can see from the curves of $P-v$ and $P-t$ that the fluid particle undergoes continuous compression−expansion processes before the HTI occurs (i.e., the part before point c). This is because the joint influences of pressure pulse and NTC phenomenon. It is noted that the compression is dominant during LTI and ITI stages. During the transition from ITI to HTI (i.e., part bc on each curve), the gas at the particle location is compressed intensively with rapidly increased pressure. Meanwhile, the heat release rate increases rapidly and reaches the maximum value when HTI occurs (i.e., point c on the $Q-t$ curve). After HTI, the pressure keeps increasing, and finally reaches its equilibrium value (see the $P-t$ curve), because the thermal explosion is achieved.

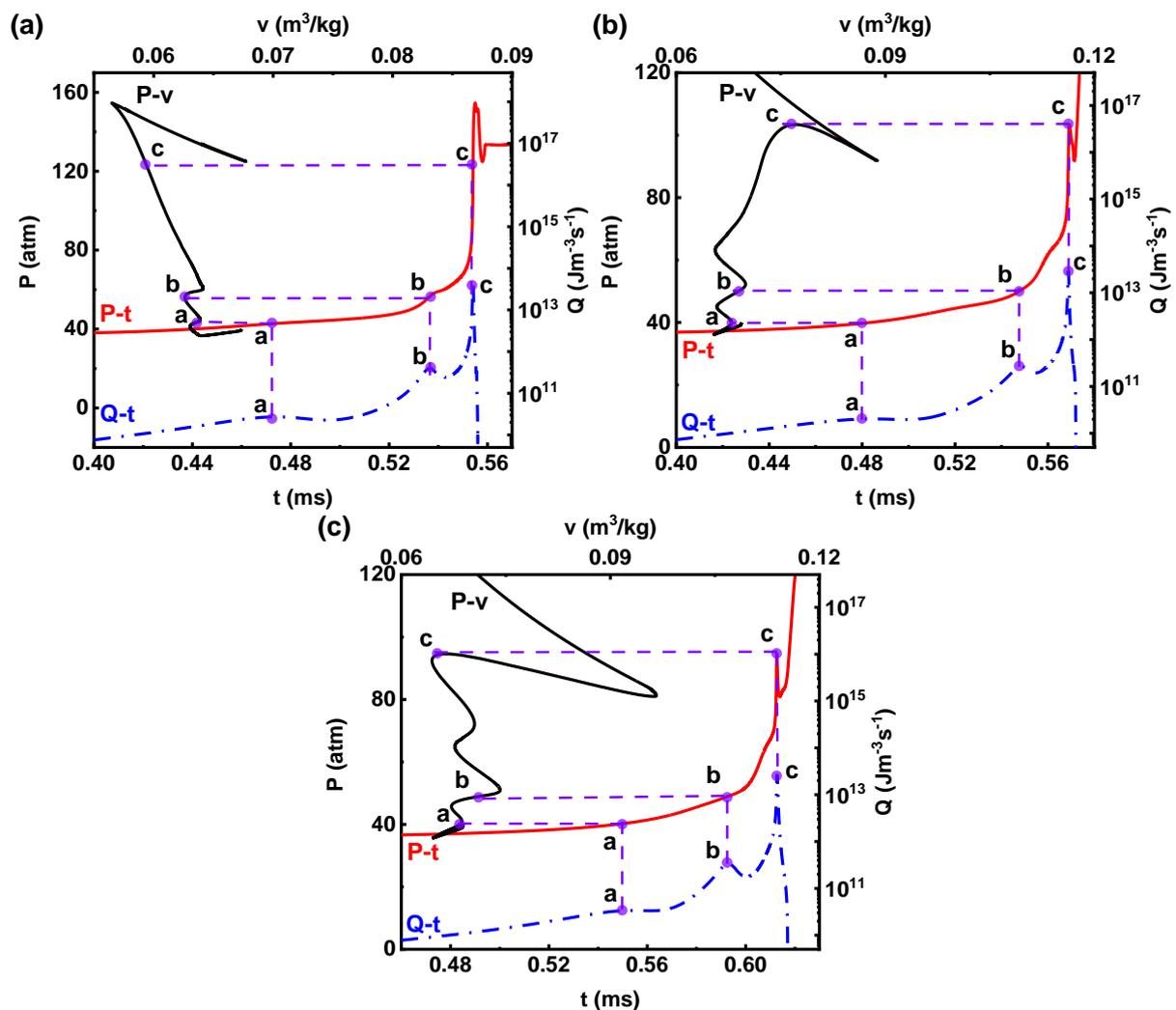

Fig. 15 Evolution of thermal states of a particle initially at midpoint of the ignition spot (i.e., $r = r_0/2$) of cases (a) A, (b) B, and (c) C. $P-v$ (pressure versus specific volume), $P-t$ (pressure versus time), and $Q-t$ (heat release rate versus time) curves are shown.

In cases B and C, continuous compression−expansion processes are also observed before the occurrence of HTI (i.e., before point c). However, the expansion is dominant during LTI and ITI stages both for case B and C, which is different from case A. During the transition from ITI to HTI (i.e., part bc), the fluid around the particle in case B is expanded, whilst in case C is compressed. This is because two opposite autoignition waves are formed within the ignition spot in case B. It is seen from Fig. 11 that the HTI firstly occurs at $r \approx 0.4r_0$. Therefore, influenced by the leftward propagating wave, the gas at the midpoint is initially expanded. Nevertheless, detonation does not develop when the pressure wave passes the particle in cases A − C. Therefore, only moderate interactions between chemical reaction and pressure wave occur.

## 5.5 Bradley's diagram for two-phase mixtures

Two parameters are used by Bradley and his co-workers [9,10] to characterize the interactions between the reaction wave and acoustic wave within the ignition spot in gaseous mixtures. The first one is the normalized temperature gradient $\xi$, which is the ratio of local sound speed to autoignition front propagation speed and measures the coupling between the local autoignition and acoustic wave caused by the heat release (or *acoustic − induction coupling* [70]). It reads

$$\xi = \frac{dT_0/dr}{(dT_0/dr)_{c,r_0/2}}. \tag{27}$$

Here $(dT_0/dr)_{c,r_0/2}$ is the critical temperature gradient, from Eq. (26). The subscript "$r_0/2$" indicates the quantity is estimated based on the initial thermochemical properties in the middle of the ignition spot. Note that $dT_0/dr$ is the initial temperature gradient within the ignition spot (see Eq. 20).

Besides, the second parameter, $\varepsilon$, is used to measure the timescale of reaction heat release relative to the residence time of the acoustic wave in the ignition spot (or

*acoustic−exothermicity coupling*). It is defined as the ratio of acoustic time to excitation time, i.e.,

$$\varepsilon = \frac{r_0/a_{r_0/2}}{\tau_e}. \tag{28}$$

where $a_{r_0/2}$ is the sound speed at the middle of the ignition spot. Here $\tau_e$ is obtained from 0D calculations based on the properties at the middle of the ignition spot.

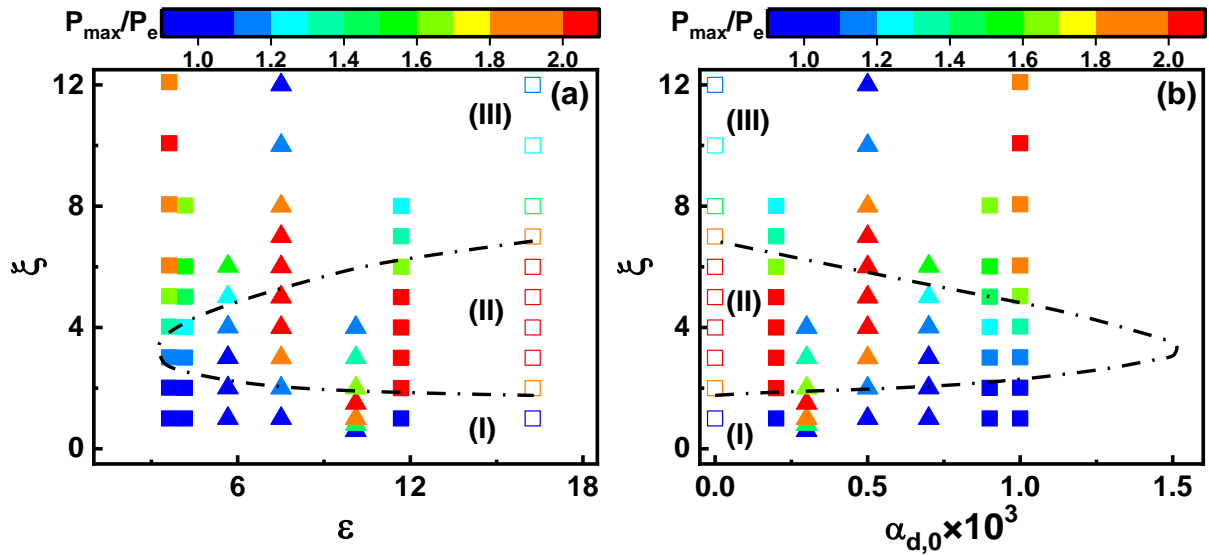

Fig. 16 Autoignition mode of stoichiometric $n$-$C_7H_{16}$/air mixtures with water droplets in (a) $\xi - \varepsilon$ and (b) $\xi - \alpha_{d,0}$ diagrams. Hollow squares: droplet-free cases initialized by hot spot; solid squares: two-phase cases initialized by hot spot; solid triangles: two-phase cases initialized by cold spot; dashed lines: detonation limits of $n$-$C_7H_{16}$/air mixtures with water vapour [38].

With the above two parameters ($\xi$ and $\varepsilon$), the autoignition modes of two-phase stoichiometric $n$-$C_7H_{16}$/air mixtures due to temperature gradient in an ignition spot are presented in Fig. 16. Note that only the autoignition waves formed at the left boundary of the ignition spot and travelling from left to right (modes d and e) are shown. The detonation limits of droplet-free mixtures diluted by $H_2O$ vapour from our previous work [38] are also plotted here, and the volume fractions for them are converted from the mole fractions of $H_2O$ vapour.

Normalized maximum pressure $P_{max}/P_e$ (with $P_{max}$ and $P_e$ being the maximum pressure from 1D calculations and equilibrium pressure from 0D constant-volume calculations, respectively) is used to identify different categories. It is seen from Fig. 16(a) that, for $\varepsilon$ = 3.6, 7.5, 10.1, 11.7, and 16.2 (corresponds to droplet-free mixtures), three categories are identified when $\xi$ increases: (I) supersonic deflagrative wave, (II) detonative wave with high maximum pressure ($P_{max}/P_e \geq 2$, red symbols in Fig. 16) and (III) subsonic deflagrative wave. This is consistent with what are found in gaseous mixtures [38]. However, the detonation regions change significantly and non-monotonically with $\varepsilon$. For example, the regional centre of detonation mode is $\xi \approx 4.5$ for $\varepsilon$ = 16.2, $\xi \approx 1.5$ for $\varepsilon$ = 10.1, and $\xi \approx 10.0$ for $\varepsilon$ = 3.6, respectively. This indicates that the effects of water droplets on $\xi$ is not monotonic, which is related to the change of $(dT_0/dr)_{c,r_0/2}$ due to ultrafine water droplet evaporation.

Furthermore, for $\varepsilon$ = 4.2 and 5.7, only supersonic deflagrative waves are found. Two opposite autoignition waves are formed when keeping increasing $\xi$ for the corresponding $\varepsilon$ (see Figs. 7b, 7c, and 10), which are not shown in Fig. 16. The distributions of the corresponding results are reversed in $\xi - \alpha_{d,0}$ in Fig. 16(b). This is because a higher droplet volume fraction $\alpha_{d,0}$ corresponds to a higher excitation time $\tau_e$, thus a lower $\varepsilon$, as shown in Eq. (28) and Fig. 5(a). Thus, one can see that, unlike the detonation limits of water droplet free mixtures (the dash-dot lines in Fig. 16), those of two-phase mixtures are not peninsular-shaped, as they are for gaseous mixtures [10,13,22–24]. Therefore, the applicability of the $\xi - \varepsilon$ diagram for a wider range of droplet-laden mixtures merits further studies, due to the significant influences of the evaporating disperse phase (e.g., water or fuel spray mists) for the thermochemical property in the ignition spot.

## 5 Conclusion

The effects of low-temperature chemistry induced by ultrafine water droplet evaporation on reaction front development from an ignition spot with temperature gradient are studied in this work. The Eulerian-Eulerian method is used to simulate the gas−liquid two-phase reactive flows and the physical model is one-dimensional spherical reactor filled with stoichiometric gaseous $n$-$C_7H_{16}$/air mixture and ultrafine water droplets (initial diameter 5 µm). The main findings are summarised below.

The results from the homogeneous autoignition in two-phase mixtures show that the dependence of ignition delay on initial gas temperature is considerably affected by the water droplet evaporation. The turnover temperatures for NTC range increase in the two-phase mixtures compared to those of the droplet-free mixtures, and also increase with droplet volume fraction. It is seen that, due to the smallness of the water droplets, they complete the evaporation and hence considerably reduce the gas temperature before the ignition occurs. Besides, only high-temperature ignition is observed when the initial droplet volume fraction is less than $5.0 \times 10^{-4}$. Beyond that, multi-stage ignitions are induced by droplet evaporation. It is also found that the excitation time increases with droplet volume fraction. Moreover, as volume fraction increases, cold or hot spot is needed to initiate a reactive front. The CEMA analysis also unveils the low-temperature chemical reactions in the gas phase chemistry induced by the ultrafine droplet evaporation.

Through the one-dimensional simulations with ignition spot, we identify three modes for the origin of the reaction front, i.e., left and right ends of ignition spot and inside it, based on the relations between gas temperature and turnover temperature for NTC range. The reaction front development corresponding to the above modes are discussed in detail. Besides, the HTI pressure wave associated with HTI ignition is more important. The predicted right-ward propagating wave speed is lower than the C−J speed of stoichiometric droplet-free $n$-$C_7H_{16}$/air mixtures. Continuous compression−expansion processes are found from $P-v$ curve, which are induced by the joint influences of pressure pulse and NTC phenomenon. Moreover, autoignition modes are summarized in $\xi - \varepsilon$ and $\xi - \alpha_{d,0}$ diagrams. The detonation regions change significantly and non-monotonically with $\varepsilon$ or $\alpha_{d,0}$. The detonation limits of two-phase mixtures are not regularly peninsular-shaped, like those for purely gaseous mixtures.

In this work, only ultrafine mono-sized water droplets are considered, which are completely gasified before ignition starts. Therefore, their direct interactions with the reaction front propagation and detonation are not present. Furthermore, the applicability of the $\xi - \varepsilon$ diagram for a wider range of droplet-laden mixtures needs to be further examined, due to the significant modulation from the evaporating disperse phase (e.g., water or fuel spray mists) for the thermochemical property in the ignition spot. These are interesting topics for our future studies.

## Acknowledgement

This work used the ASPIRE 1 Cluster from National Supercomputing Centre, Singapore (https://www.nscc.sg/). ZY is supported by the NUS Research Scholarship Budget (Grant Nos. C-261-000-207-532 and C-261-000-005-001). Wantong Wu and Qiang Li are thanked for helpful discussions about CEMA calculations and evaporation model implementations, respectively.

**Appendix A. Validation of droplet heating and evaporation models**

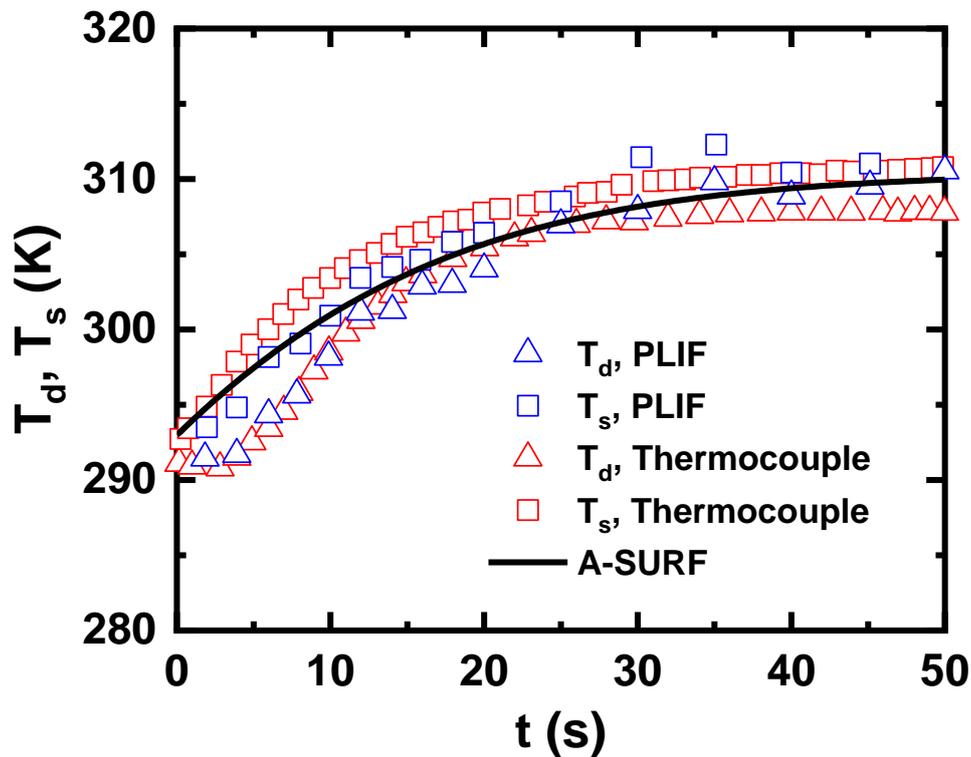

Fig. A1 Time history of the droplet temperature. Experimental data: Volkov and Strizhak [71].

Figure A1 shows the time history of temperature of single water droplet. In the experiment by Volkov and Strizhak [71], one water droplet is placed in the air with temperature of 373 K and velocity of about 3 m/s. Two measurement techniques are used to determine the surface and internal droplet temperatures $T_s$ and $T_d$ (both included in Fig. A1), namely Planar Laser Induced Fluorescence (PLIF) and thermocouple. The droplet is not exactly spherical, and its volume is 10 μL (the corresponding nominal diameter is 2.67 mm) [71]. In the simulation, we assume that the temperature inside the droplet is uniform, and the initial droplet diameter equals to the nominal diameter. It is seen from Fig. A1 that good agreement can be achieved about the overall evolutions of the droplet temperature and the equilibrium temperature around $t = 50$ s.

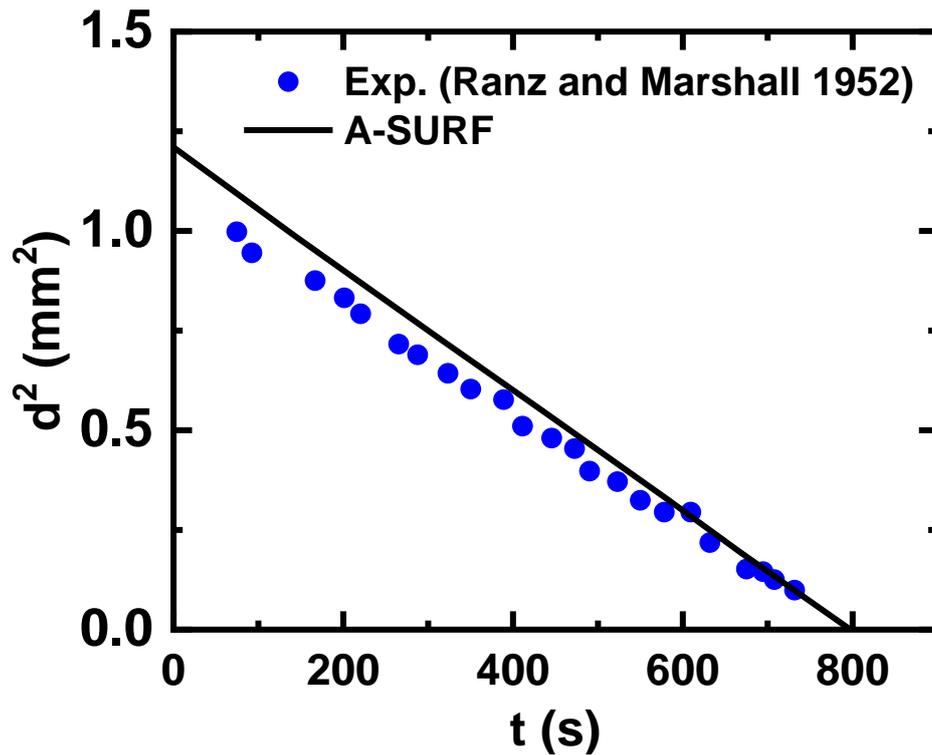

Fig. A2 Time history of the square of droplet diameter. Experimental data: Ranz and Marshall [53].

Figure A2 further compares the diameter evolution of an evaporating water droplet against the experimental data [53]. The initial diameter and temperature of the water droplet are 1.1 mm and 282 K, respectively. The temperature of the ambient gas is 298 K. Excellent agreement is found between the present simulations and the measured data. Besides, the slope (i.e., evaporation coefficient) computed by A-SURF is $-1.51 \times 10^{-3}$ mm$^2$/s, close to that from experimental data ($-1.37 \times 10^{-3}$ mm$^2$/s). In general, the evaporation model in A-SURF can accurately predict the droplet evaporation in terms of the evaporation coefficient and the diameter decaying history.